\documentclass[fleqn,10pt]{wlscirep}
\usepackage[utf8]{inputenc}
\usepackage[T1]{fontenc}

\usepackage{amsmath}
\usepackage{amsthm}
\usepackage{algorithm2e}
\SetKwRepeat{Do}{do}{while}%
\usepackage{multirow}
\title{Multi-Qubit Correction for Quantum Annealers}

\author[1]{Ramin Ayanzadeh}
\author[2]{John Dorband}
\author[2]{Milton Halem}
\author[2]{Tim Finin}

\affil[1]{College of Computing, Georgia Institute of Technology, Atlanta, GA 30332, United States}
\affil[2]{Computer Science and Electrical Engineering, University of Maryland, Baltimore County, Baltimore, MD 21250, United States}
\affil[*]{ayanzadeh@gatech.edu}

\keywords{Adiabatic Quantum Computing, Error-Correction, Noisy Intermediate-Scale Quantum (NISQ) Computers, Optimization, Quantum Annealing}

\begin{abstract}
We present \emph{multi-qubit correction} (MQC) as a novel postprocessing method for quantum annealers that views the evolution in an open-system as a Gibbs sampler and reduces a set of excited states to a new synthetic state with lower energy value. 
After sampling from the ground state of a given (Ising) Hamiltonian, MQC compares pairs of excited states to recognize virtual tunnels—i.e., a group of qubits that changing their states simultaneously can result in a new state with lower energy value—and successively converges to the ground state. 
Experimental results using D-Wave 2000Q quantum annealers demonstrate that MQC finds samples with notably lower energy values and improves the reproducibility of results when compared to recent hardware/software advances in the realm of quantum annealing, such as spin-reversal transforms, classical postprocessing techniques, and increased inter-sample delay between successive measurements.
\end{abstract}

\begin{document}

\flushbottom
\maketitle
\thispagestyle{empty}

 \section{Introduction}
Quantum annealing is a meta-heuristic for addressing discrete (or combinatorial) optimization problems that are intractable in the realm of classical computing. 
While simulated annealing (a.k.a. thermal or classical annealing) uses adjustable thermal fluctuations to jump over the energy mountains, quantum annealing applies adjustable quantum fluctuations for tunneling through the (narrow-enough) energy barriers \cite{amara1993global,finnila1994quantum,kadowaki1998quantum,das2008colloquium,ohzeki2011quantum,nishimori2017exponential}. 
Quantum annealers are a special case of the adiabatic quantum computers (i.e., stoquastic open-system) that provide a hardware implementation for finding the ground state (or minimum energy configuration) of (Ising) Hamiltonians \cite{albash2018adiabatic,ayanzadeh2020reinforcement}.
To solve a problem using the quantum annealers, therefore, we must define a Hamiltonian whose ground state represents the optimum solution of the original problem of interest \cite{lucas2014ising,mcgeoch2020theory,ayanzadeh2020leveraging}. 

We can form an Ising Hamiltonian whose ground state represents the optimum solution of any given problem of interest—which can be nontrivial in many real-world applications \cite{ayanzadeh2019quantum,ayanzadeh2020leveraging}. 
In practice, however, executing the corresponding quantum machine instruction (QMI) on a physical quantum annealer does not guarantee achieving the global optimum \cite{albash2018adiabatic,ayanzadeh2019quantum_assisted,ayanzadeh2020reinforcement}. 
In addition to thermal noise and diabatic transitions \cite{albash2018adiabatic}, examples of control error sources include sparse connectivity between qubits \cite{cai2014practical,vinci2015quantum}, confined annealing schedule \cite{nishimori2017exponential}, coefficients’ range and precision limitations \cite{pudenz2015quantum,dorband2018extending}, and noise and decoherence \cite{lidar2008towards,deng2013decoherence,pudenz2014error,gardas2018defects,gardas2018quantum}. These error sources lower the quality of results, i.e., the energy value of the drawn samples is higher than the energy value of the ground state \cite{king2014algorithm}. 

Modifying some aspects of the Hamiltonian by adapting (and better selecting) initial and final Hamiltonians, optimizing the schedule/path function, adding a catalyst Hamiltonian (i.e., a Hamiltonian that is present only in intermediate time), or adding non-stoquastic term to the Hamiltonian can circumvent certain drawbacks of the adiabatic quantum computers \cite{albash2018adiabatic}. 
Nevertheless, the majority of these techniques are mainly suitable for closed systems, or current generations of the physical quantum annealers cannot (fully) accommodate them. 
Acknowledging that adiabatic quantum computing has some inherent resistance to noise and decoherence, we need error correction and mitigation mechanisms for ensuring the scalability of adiabatic quantum computers just as we do with other quantum information processing models \cite{jordan2006error,sarovar2013error,mizel2014fault,vinci2016nested}. 
In spite of several error correction proposals for adiabatic quantum computing and quantum annealing \cite{jordan2006error,lidar2008towards,sarovar2013error,young2013adiabatic,young2013error,ganti2014family,mizel2014fault,pudenz2014error,bookatz2015error,pudenz2015quantum,vinci2015quantum,matsuura2016mean,mishra2016performance,vinci2016nested,matsuura2019nested}, 
an accuracy-threshold theorem for adiabatic quantum computing, unlike its gate model counterpart \cite{aliferis2005quantum}, remains elusive \cite{pudenz2014error,vinci2016nested}. 
Besides, most error correction schemes (e.g., nested quantum annealing correction method \cite{vinci2016nested,matsuura2019nested}) utilize multiple physical qubits for coding every qubit that notably reduces the capacity of current quantum annealers. 

From an application perspective, problem-solving with a physical quantum annealer has two drawbacks: (1) quantum annealers can yield excited states rather than the ground state of the given Hamiltonian, and (2) the results/samples attained by the physical quantum annealers are not reproducible over time.
Applying classical postprocessing techniques to (raw) samples attained by the physical quantum annealers can mitigate these drawbacks to some extent. 
For example, one may apply evolutionary algorithms (e.g., genetic algorithms) or swarm intelligence techniques (such as particle swarm optimization) as a postprocessing method for quantum annealers \cite{engelbrecht2007computational}. 
Nevertheless, these heuristics and meta-heuristics are stochastic techniques that can result in a different solution in each try and lessen the robustness of results in the application domain. 
Moreover, most of these techniques require hyper-parameter optimization—e.g., the number of iterations, the probability of crossover and mutation, and selection method in evolutionary algorithms—that can significantly impact the quality of ultimate results \cite{shahamatnia2011adaptive}. 
To this end, we devise a lightweight and deterministic postprocessing method that can improve the quality and robustness of results attained by physical quantum annealers. 
It is worth highlighting that, in this study, we did not aim to demonstrate quantum speed-up.

We view quantum annealers as a Gibbs sampler that allows diabatic transitions in an open system and present \emph{multi-qubit correction} (MQC) as a novel postprocessing technique for quantum annealers. 
Unlike most studies that try to identify different types of errors and mitigate/correct them, we try to recognize the pattern(s) among the incorrect observations (or lower quality samples compared to the ground state of the given Hamiltonian) and leverage it to achieve a better solution.
Measurement is the most error-prone operation on superconducting noisy intermediate-scale quantum (NISQ) \cite{preskillNISQ} machines \cite{tannu2019mitigating}—here, we look at the quantum annealers as the NISQ model of adiabatic quantum computers. 
Therefore, in this study, we mainly focus on bit-flips that mostly occur due to measurement errors on superconducting quantum devices. 
MQC compares pairs of samples and recognizes groups of qubits that we can flip their values simultaneously, and tunnels through these groups to converge to a (notably) better solution. 
Our experiments using the D-Wave 2000Q quantum annealers show that MQC utilizes fewer samples and finds better solutions compared to recent software/hardware advances in the realm of quantum annealing.

\section{Results}
Quantum annealers, like the quantum processing units (QPU) by D-Wave Systems, are single-instruction (quantum) computing machines that can only sample from the ground state of the following problem Hamiltonian (denoted by $H_p$): 
\begin{equation}
	\label{eqn:ising_energy}
	H_p := E_{\text{Ising}} \left( \mathbf{z} \right) = \sum_{i=1}^{N}{\mathbf{h}_i \mathbf{z}_i} + \sum_{i=1}^{N}{\sum_{j=i+1}^{N}{J_{ij} \mathbf{z}_i \mathbf{z}_j}},
\end{equation}
where $N$ denotes the number of quantum bits (qubits), spin variables $\mathbf{z} \in \{-1,+1\}^N$, and $\mathbf{h}$ and ${J}$ represent local fields  and couplers, respectively \cite{johnson2011quantum,mcgeoch2020theory,ayanzadeh2020leveraging}. 
Quantum annealers can efficiently recognize the region of the ground state(s) of the given Hamiltonians; however, they generally fail to get to the global minimum, regardless of how close they are to the ground state.
In other words, unlike classical annealing that always converges to a local (or sometimes the global) optimum, quantum annealers generally yield an excited state(s) that are not necessarily a local optimum \cite{dorband2018method,ayanzadeh2019quantum_assisted,ayanzadeh2020reinforcement}. 
Hence, we can expect that applying optimization heuristics and meta-heuristics on samples attained by a quantum annealer to result in new (or synthetic) sample(s) with lower energy value, specifically on systems with glassy landscapes \cite{das2008colloquium}. 
In this section, we start with a local optimization heuristic, called \emph{single-qubit correction} (SQC), and then extend it to introduce \emph{multi-qubit correction} (MQC) scheme for mitigating errors in quantum annealers \cite{dorband2018method}. 

\subsection{Single-Qubit Correction}
Measuring qubits after the annealing process results in an eigenstate that is not necessarily the ground state or even a local optimum of the given problem Hamiltonian \cite{das2008colloquium}. 
Owing to hardware limitations such as limited precision of coefficients, the Hamiltonian that is minimized on a physical quantum annealer can be (slightly) different from the problem Hamiltonian of interest (in application domain); therefore, the evolution is not guaranteed to result in an eigenstate of the problem Hamiltonian \cite{albash2018adiabatic}. 
We start by adopting a hill-climbing \cite{engelbrecht2007computational,russell2016artificial} algorithm to present a postprocessing approach for quantum annealers, called \emph{single-qubit correction} (SQC). 
SQC has a zero-temperature simulated annealing scheme that can relax a raw sample to a new/synthetic sample with a lower energy value. 
In each iteration of SQC, we individually toggle the value of every qubit and keep all of the changes that result in a state with a lower energy value. 
Algorithm \ref{algo:SQC} shows how SQC exploits the neighborhood of an excited state, denoted by $\mathbf{z}.$ 

\begin{algorithm}[!ht]
	\DontPrintSemicolon 
	\KwIn{$\mathbf{z}, \mathbf{h}$ and $J$}
	\KwOut{$\mathbf{z}$}
	$N \gets \lvert\mathbf{z}\rvert$\;		
	$lowestEnergy \gets E_{\text{Ising}}(\mathbf{z}, \mathbf{h}, J)$\;
	$terminate \gets \text{True}$\;		
	\Do{$\mathrm{\mathbf{not\;}} terminate $}{	
		\For{$i \gets 1$ \KwTo $N$}{
			$\mathbf{z}_i \gets -\mathbf{z}_i$ \tcp*{Flip bit and check its influence}
			\eIf{$lowestEnergy > E_{\text{Ising}}(\mathbf{z}, \mathbf{h}, J)$}{ 
				$lowestEnergy \gets E_{\text{Ising}}(\mathbf{z}, \mathbf{h}, J)$\;
				$terminate \gets \text{False}$\;
			}{
				$\mathbf{z}_i \gets -\mathbf{z}_i$ \tcp*{Flip bit back since it did not result in a sample with lower energy}
			}
		}
	}	
	\Return{$\mathbf{z}$}\;
	\caption{
Single-qubit correction (SQC) heuristic for exploiting the neighborhood of excited states to find a sample with lower energy value.
}
	\label{algo:SQC}
\end{algorithm}

\begin{figure}[!ht]
	\begin{center}
	\includegraphics[scale=1]{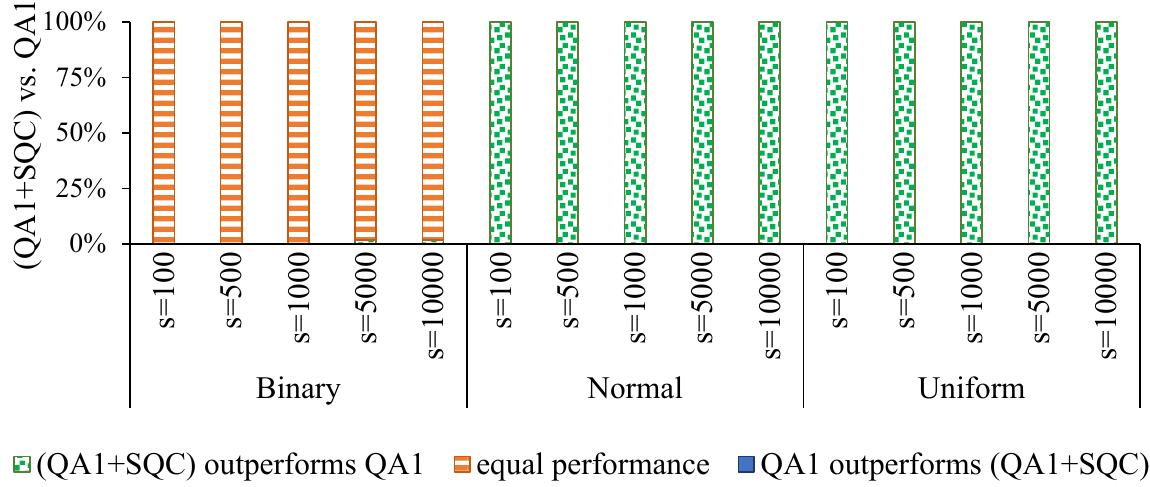} 
	\end{center}
	\vspace{-0.6cm}
	\caption{
Performance comparison between quantum annealing ($\text{QA}^1)$ and applying SQC to raw samples attained by quantum annealers ($\text{QA}^1+SQC)$ where $s$ denotes the number of samples. 
}
	\label{fig:benchmarking_QA1_QA1+SQC}
\end{figure}

From an optimization point of view, SQC is very likely to result in a meta-stable state (i.e., SQC is sensitive to the input sample) when the input sample is not a local optimum, which is trivial for systems with the glassy landscape. 
To solve a problem on a quantum annealer, we generally draw many samples, e.g., up to 10,000 samples/reads per QMI on a D-Wave quantum annealer. 
Therefore, in practice, SQC explores a broader area, i.e., the neighborhood of all excited states attained by the quantum annealers.
Figure \ref{fig:benchmarking_QA1_QA1+SQC} illustrates the impact of applying SQC to raw samples drawn by the D-Wave quantum annealers. 
For uniform and normal benchmark problems, SQC always finds a better sample, and the corresponding $p-$value is $8.8818 \times 10^{-16}$ which indicates that results are statistically significant (i.e., $p<0.05)$. 
However, for Binary problems, both SQC and the baseline method demonstrated a similar performance (i.e., results are not statistically significant). 
For more information about the benchmarking, see the section Method.

It is worth highlighting that we do not propose SQC as a postprocessing method for quantum annealers since more efficient techniques,  such as simulated annealing, can outperform SQC in terms of finding samples with lower energy values. 
Indeed, we will extend SQC to introduce a novel postprocessing approach that can notably improve the quality and reproducibility of results attained by the quantum annealers. 

\subsection{Multi-Qubit Correction}
SQC is an optimization heuristic that (1) neglects the interactions between spins of the given problem Hamiltonian; and (2) entirely depends on one excited state as its initial state.
Instead of processing one qubit at a time on samples individually, we introduce \emph{multi-qubit correction} (MQC) method that treats groups of qubits as tunnels (or units) and compares pairs of samples to find these tunnels. 
In this study, the term “tunnel” is analogous to the concept of quantum tunnels \cite{denchev2016computational} and refers to a group of qubits that a quantum annealer simultaneously flips their values to change an excited state to a sample with a lower energy value. 

Let $\mathbf{h}$ and ${J}$ denote linear and quadratic coefficients of a problem Hamiltonian that we aim to find its ground state, respectively.
Also, let $\mathbf{z}^{1}$ and $\mathbf{z}^{2}$ be two (excited) states, with ${N}$ spin variables, attained by a quantum annealer. 
We define two sets of qubits as 
\begin{equation}
	\label{eqn:MQC_S}
	S=\{ \mathbf{z_i} | \mathbf{z}_{i}^{1} = \mathbf{z}_{i}^{2}\}
\end{equation}
and
\begin{equation}
	\label{eqn:MQC_D}
	D=\{ \mathbf{z_i} | \mathbf{z}_{i}^{1} \neq \mathbf{z}_{i}^{2}\}
\end{equation}
where $i=1, 2, \dots, N.$
The set ${D}$ is a tunnel that represents the transformation of sample $\mathbf{z}^{1}$ to sample $\mathbf{z}^{2}$, and vice versa. 
From a problem-solving viewpoint, finding $D$ is rather useless. 
However, we can use $D$ to find sub-tunnels that may reduce $\mathbf{z}^{1}$ (or $\mathbf{z}^{2})$ into a new sample with a lower energy value. 
A sub-tunnel of ${D},$ denoted by ${T},$ is a subset of ${D}$ where 
\begin{equation}
	J_{ij} = 
	\begin{cases}
		\mathbb{R}_{\neq0}	&	\mathbf{z}_i, \mathbf{z}_j \in T;\\
		0	&	\mathbf{z}_i \;\text{or}\; \mathbf{z}_j \notin T.
	\end{cases}
\end{equation}
In this sense, ${T}$ is the closure of a set of qubits connected transitively to each other but not connected to other qubits in ${D}.$ 
Hence, we can represent ${D}$ as a partition of sub-tunnels as 
\[
	D=\bigcup_k{T^k}
\] 
where
\[
	T^i \cap T^j = \{\}, \quad\quad \forall i,j.
\]
We define the influence (or energy contribution) of $T^k$ to $\mathbf{z}^{1}$ as follows:
\begin{equation}
	\label{eqn:MQC_influence_subtunnel}
	I^k_{\mathbf{z}^{1}} = {\sum\limits_{i \in T^k} \mathbf{h}_i\mathbf{z}^{1}_i + \sum\limits_{i \in T^k} \sum\limits_{j \in S} J_{ij} \mathbf{z}^{1}_i \mathbf{z}^{1}_j }.
\end{equation}
Note that we have omitted the term 
\[
	\sum\limits_{i,j \in T^k} J_{ij} \mathbf{z}^{1}_i \mathbf{z}^{1}_j
\]
from Eq. \eqref{eqn:MQC_influence_subtunnel} since flipping values of all qubits in $T^k$ does not effect $I^k_{\mathbf{z}^{1}}.$
Finally, we reduce $\mathbf{z}^{1}$ to a new sample, denoted by $\mathbf{z}^*,$ via flipping the qubit values of all sub-tunnels that have a positive influence (or energy contribution) value.
Note that $I^k_{\mathbf{z}^{1}} = -I^k_{\mathbf{z}^{2}};$ thus, applying the abovementioned process on $\mathbf{z}^{2}$ will result in a new sample that is identical to $\mathbf{z}^*.$
Algorithm \ref{algo:MQC_reduce} shows how we reduce two input samples, denoted by $\mathbf{z}^1$ and $\mathbf{z}^2$, to a new sample (denoted by $\mathbf{z}^*)$ whose energy value is guaranteed to be less than or equal to energy values of $\mathbf{z}^1$ or $\mathbf{z}^2.$

\begin{algorithm}[!ht]
	\DontPrintSemicolon
	\SetKwFunction{FReduce}{Reduce}
	\SetKwProg{Fn}{Function}{:}{}
	\Fn{\FReduce{$\mathbf{z}^1$, $\mathbf{z}^2$, $\mathbf{h}$, $J$}}{
		$\mathbf{z}^* \gets \mathbf{z}^1$\;
	      $N \gets \lvert\mathbf{z}^*\rvert$\;
		$S \gets \{\}$\;
		$D \gets \{\}$\;
		\For{$i \gets 1$ \KwTo $N$}{
			\eIf{$\mathbf{z}^1_i = \mathbf{z}^2_i $}{
				$S \gets S \cup \{i\}$ \tcp*{Identical bits between $\mathbf{Z}^1$ and $\mathbf{Z}^2$}
			} {
				$D \gets D \cup \{i\}$ \tcp*{Different bits between $\mathbf{Z}^1$ and $\mathbf{Z}^2$}
			}
		}
		$Adj \gets [\,]$ \tcp*{Adjacency of bits in $D$ based on $J$}
		\For{$i \in D$}{
			$Adj_i =\{\}$\;
		}
		\For{$i,j \in J$}{
			\If{$J_{ij}\neq0$ $\&$ $i \in D$ $\&$ $j \in D$}{
				$Adj_i \gets Adj_i \cup \{j\}$\;
				$Adj_j \gets Adj_j \cup \{i\}$\;
			}
		}
		\SetKwFunction{Fcc}{ConnectedComponents}
		$T \gets \Fcc{$Adj$}{}$\tcp*{Sub-tunnels (groups of isolated bits in $D)$}
		\For{$k \gets 1$ \KwTo $|T|$}{
			$I^k_{\mathbf{z}^1} = \sum\limits_{i \in T^k} \mathbf{h}_i \mathbf{z}^1_i + \sum\limits_{i \in T^k} \sum\limits_{j \in S} J_{ij} \mathbf{z}^1_i \mathbf{z}^1_j$ \tcp*{Influence of sub-tunnels on $E_{\text{Ising}}$}
			\If{$ I^k_{\mathbf{z}^1} > 0$}{
				\For{$l \in T^k$}{
					$\mathbf{z}^*_l \gets -\mathbf{z}^*_l$ \tcp*{Flip all bits of sub-tunnels with positive influence}
				}

			}
		}
	}
	\Return{$\mathbf{z}^*$}\;
	\caption{Reducing two input samples $\mathbf{z}^1$ and $\mathbf{z}^2$ to a new sample ($\mathbf{z}^*)$ with lower energy value based on virtual tunnels.}
	\label{algo:MQC_reduce}
\end{algorithm}

The Reduce procedure presented in this paper is analogous to the crossover operation in evolutionary algorithms that acts on two potential solutions, known as parent chromosomes, and yields new solution(s), known as offspring \cite{engelbrecht2007computational,shahamatnia2011adaptive}. 
The Reduce procedure, shown in Algorithm \ref{algo:MQC_reduce}, is the extended version of SQC method, shown in Algorithm \ref{algo:SQC}, that acts on a group of qubits simultaneously rather than flipping the values of qubits individually. 
Besides, the Reduce procedure acts on two excited states and is less sensitive to a single initial point when compared to SQC that depends entirely on a single excited state. 

When we employ physical quantum annealers, we generally request many samples/reads, i.e., repeating the annealing process with different initial eigenstates to improve the probability of achieving the ground state of the given Hamiltonian.
Algorithm \ref{algo:MQC} illustrates the \emph{multi-qubit correction} (MQC) method that receives a sample set as input and tries to iteratively reduce it to a new sample with a lower energy value. 
For a sample set with ${n}$ samples/reads, MQC takes $\log_2(n)$ steps, and in each iteration, the size of the sample set is divided by two. 

\begin{algorithm}[!ht]
	\DontPrintSemicolon 
	\SetKwFunction{Reduce}{Reduce}
	\SetKwFunction{Pop}{Pop}
	\SetKwFunction{Append}{Append}
	\KwIn{$Z=\left[\mathbf{z}^1, \mathbf{z}^2, \dots, \mathbf{z}^n\right], \mathbf{h}$ and $J$}
	\KwOut{$\mathbf{z}^*$}
	\While{$|Z| >1$}{
		$\hat{Z} \gets Z$\;
		$Z \gets \{\}$\;
		\While{$|\hat{Z}| >1$}{
			$\mathbf{z}^A \gets \hat{Z}.\Pop{}{}$\;
			$\mathbf{z}^B \gets \hat{Z}.\Pop{}{}$\;
			$\mathbf{z}^{AB} \gets \Reduce{$\mathbf{z}^A, \mathbf{z}^B, \mathbf{h}, J$}{}$\;
			$Z.\Append{$\mathbf{z}^{AB}$}{}$\;
		}
		\If{$|\hat{Z}| >0$}{
			$Z.\Append{$\hat{Z}.\Pop{}{}$}{}$\;
		}
	}
	$\mathbf{z}^* \gets Z.\Pop{}{}$\;	
	\caption{Multi-qubit correction (MQC) method for reducing a set of samples to a new sample whose energy value is less than all input samples or (in the worst case) is equal to the lowest energy value of input samples.}
	\label{algo:MQC}
\end{algorithm}

\begin{figure}[!ht]
	\begin{center}
	\includegraphics[scale=1]{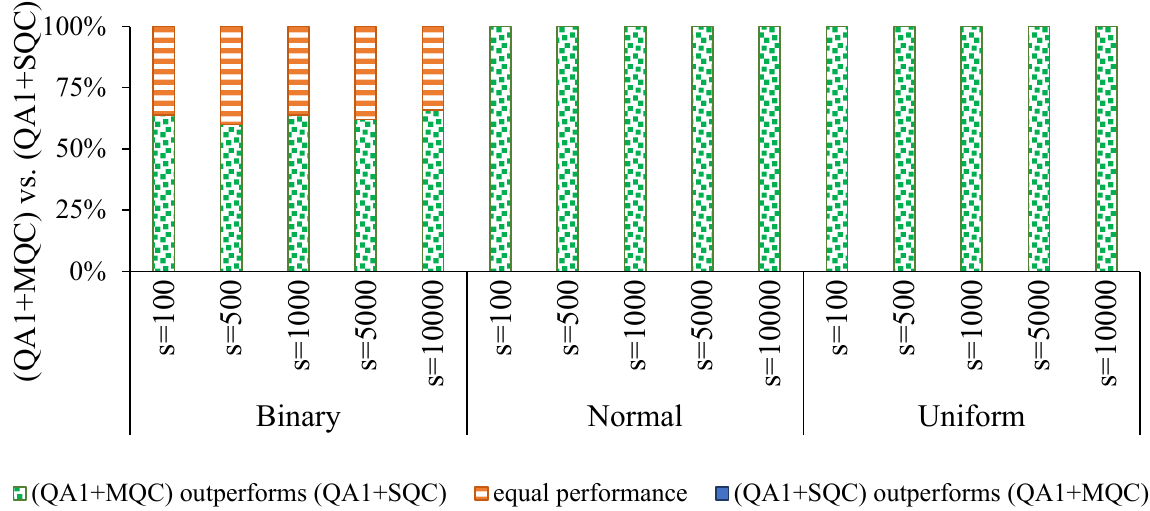} 
	\end{center}
	\vspace{-0.6cm}
	\caption{
	Performance comparison between applying MQC and SQC to raw samples attained by quantum annealers, denoted by $\text{QA}^1+MQC$ and $\text{QA}^1+SQC,$ respectively $(s$ denotes the number of samples). 
	}
	\label{fig:benchmarking_QA1+SQC_QA1+MQC}
\end{figure}

Figure \ref{fig:benchmarking_QA1+SQC_QA1+MQC} shows that for normal and uniform benchmark problems MQC always outperforms SQC (i.e., finds a better sample with lower energy value), and the corresponding  $p-$value is $8.8818 \times 10^{-16}.$  
For Binary benchmark problems, the success rate of MQC (i.e., MQC outperforms SQC) ranges from 60\% to 66\%, and the corresponding $p-$values ranges from $9.3132 \times 10^{-10}$ to $1.1641 \times 10^{-10}$ which indicate that results are statistically significant. 
We remark that SQC was not able to outperform MQC in any arrangement. 

\begin{figure}[!ht]
	\begin{center}
	\includegraphics[scale=1]{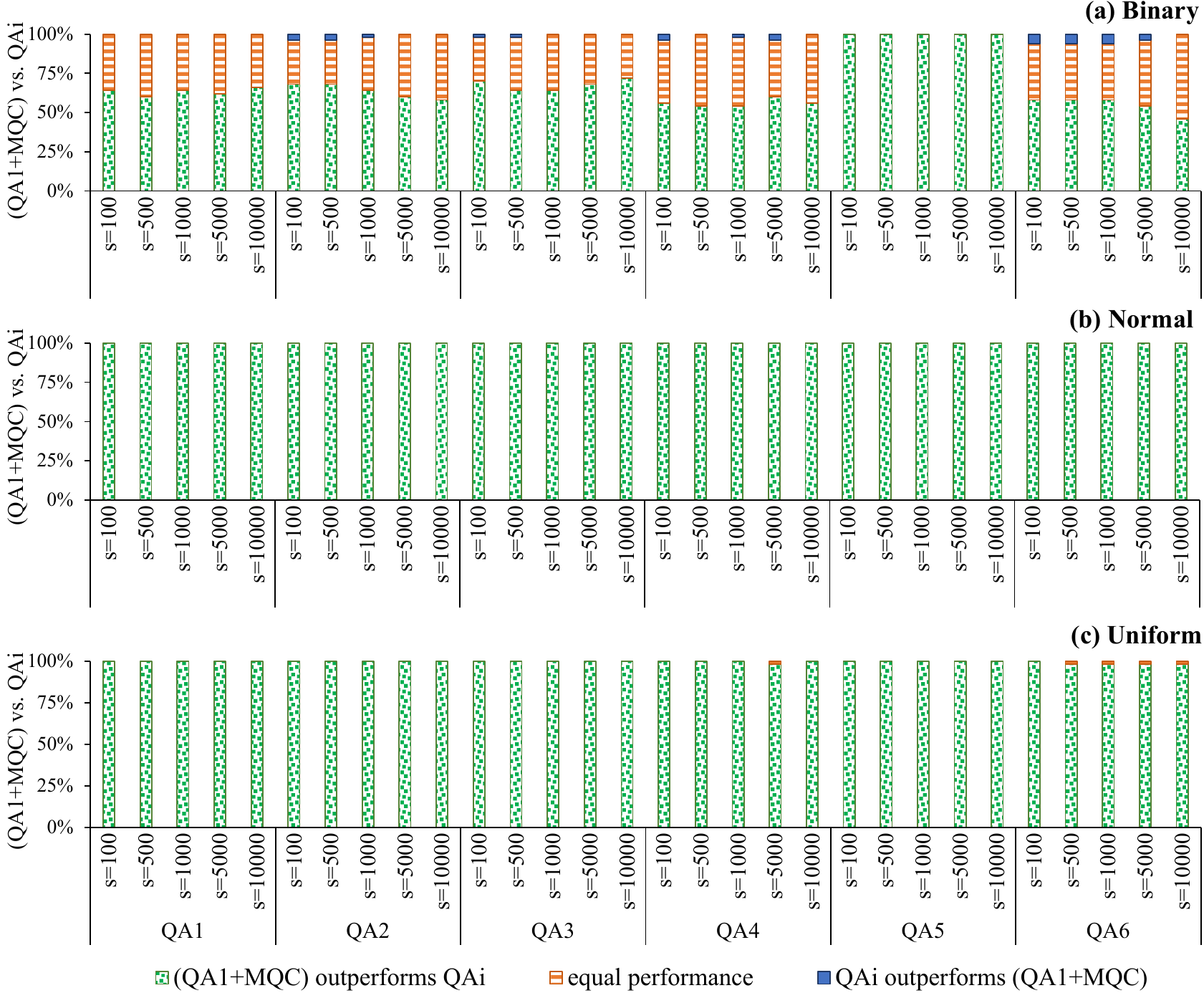} 
	\end{center}
	\vspace{-0.6cm}
	\caption{
	Comparing the performance of applying MQC to raw samples attained by a quantum annealer ($\text{QA}^1+MQC)$ with recent software/hardware advances—namely quantum annealing with five spin-reversal transforms $(\text{QA}^2),$ quantum annealing with longer inter-sample delay $(\text{QA}^3),$ quantum annealing with optimization postprocessing $(\text{QA}^4),$ quantum annealing with sampling postprocessing $(\text{QA}^5),$ and quantum annealing with five spin-reversal transforms, longer preparation time and optimization postprocessing $(\text{QA}^6)$ where $s$ denotes the number of samples. 
	}
	\label{fig:benchmarking_QAi_QA1+MQC}
\end{figure}

Figure \ref{fig:benchmarking_QAi_QA1+MQC} demonstrates that applying MQC on raw samples, attained by the D-Wave quantum processors in sampling from the ground state of benchmark Ising models, outperforms recent software/hardware advances in the field of quantum annealing, such as spin-reversal transforms, optimization, and sampling postprocessing methods, and increased inter-sample delay between successive measurements. 
For normal benchmark problems, MQC always outperforms all baselines (i.e., finds a better sample with lower energy value), and the corresponding  $p-$value is $8.8818 \times 10^{-16}.$  
For uniform benchmark problems, the success rate of MQC (i.e., MQC outperforms the baseline method) ranges from 98\% to 100\%, and the corresponding $p-$values ranges from 
$1.7764 \times 10^{-15}$ to $8.8818 \times 10^{-16}.$ 
For Binary benchmark problems, the success rate of MQC (i.e., MQC outperforms the baseline method) ranges from 46\% to 100\%, and the corresponding $p-$values ranges from 
$1.1921 \times 10^{-7}$ to $8.8818 \times 10^{-16}.$ 
We remark that none of the baseline methods was not able to outperform MQC in any arrangement, and results are statistically significant (i.e., $p<0.05)$ in all arrangements. 
These results explain that MQC requires notably fewer samples to visit the ground state of the given Hamiltonian with a high enough probability compared to recent software and hardware advances in the realm of quantum annealing. 
For more information about the benchmarking, see the section Method.

From an application perspective, problem-solving with a physical quantum annealer has two drawbacks: (1) quantum annealers can yield excited states rather than the ground state of the given Hamiltonian, and (2) the results/samples attained by the physical quantum annealers are not reproducible over time. 
According to Anderson localization \cite{altshuler2010anderson}, as an example, the energy gap between the ground and first excited states is shrunk close to the end of the annealing. 
The landscape of glassy Hamiltonians generally includes many excited states, and a physical quantum annealer is likely to relax to one of these excited states.

In conclusion, problem-solving with quantum annealers results in a distribution of (potential) ground states, and the variance of the corresponding energy values is large enough to lessen the reproducibility of results.
We repeated the aforementioned methods 50 times, and Table \ref{tbl:benchmarking_robustness} includes the variance of the energy values of these repeated experiments. 
Figure \ref{fig:benchmarking_QAi_QA1+MQC} explains that applying the sampling postprocessing on raw samples attained by the D-Wave quantum annealers ($\text{QA}^5$) can significantly lower the quality of results. 
In other words, applying the sampling postprocessing method can result in a sample with a higher energy value. 
On this basis, we omitted $\text{QA}^5$ from Table \ref{tbl:benchmarking_robustness}. 
These results reveal that applying MQC on (raw) samples attained by the D-Wave quantum processors can notably improve the reproducibility of results. 

\begin{table}[!ht]
	\begin{center}
		\caption{
	Comparing the robustness (or reproducibility of results) of applying MQC to raw samples attained by a quantum annealer ($\text{QA}^1+MQC)$ with recent software/hardware advances—namely quantum annealing with five spin-reversal transforms $(\text{QA}^2),$ quantum annealing with longer inter-sample delay $(\text{QA}^3),$ quantum annealing with optimization postprocessing $(\text{QA}^4),$ and quantum annealing with five spin-reversal transforms, longer preparation time and optimization postprocessing $(\text{QA}^6).$ Each element represents the variance of energy values from repeating the corresponding method 50 times.
}
		\label{tbl:benchmarking_robustness}
		\begin{tabular}{cccccccc}
			Coefficients		&	Samples	&	$\mathrm{QA^1}$	&	$\mathrm{QA^2}$	&	$\mathrm{QA^3}$	&	$\mathrm{QA^4}$	&	$\mathrm{QA^6}$	&	$\mathrm{QA^1+MQC}$\\
			\hline
			\multirow{4}{*}{Binary}
				&	100	&	1.3696	&	2.8304	&	1.4544	&	2.7264	&	5.2416	&	0.9984\\
				&	200	&	0.9936	&	3.4576	&	0.9664	&	1.4656	&	8.1936	&	0.8704\\
				&	500	&	0.7696	&	2.3936	&	0.9104	&	1.3584	&	3.9184	&	0.6400\\
				&	1000	&	0.4816	&	1.9584	&	0.9984	&	0.8464	&	3.3936	&	0.2944\\
			\hline
			\multirow{4}{*}{Uniform}
				&	100	&	1.6558	&	1.1323	&	0.6401	&	0.6606	&	1.1116	&	0.0704\\
				&	200	&	0.8674	&	0.9988	&	0.7189	&	0.4792	&	1.2079	&	0.0171\\
				&	500	&	0.5471	&	0.9898	&	0.5596	&	0.4471	&	0.7513	&	0.0092\\
				&	1000	&	0.6006	&	0.5860	&	0.5792	&	0.2757	&	0.7606	&	0.0049\\
			\hline
			\multirow{4}{*}{Normal}
				&	100	&	5.9365	&	4.0989	&	2.5668	&	0.4009	&	1.0033	&	0.0563\\
				&	200	&	3.8940	&	3.5666	&	2.4667	&	0.3343	&	0.5284	&	0.0071\\
				&	500	&	2.8559	&	2.0883	&	2.0424	&	0.4017	&	0.4056	&	0.0000\\
				&	1000	&	2.7226	&	2.3799	&	1.5881	&	0.2467	&	0.2899	&	0.0000\\
			\hline

		\end{tabular}
	\end{center}
\end{table}

\subsection{Randomized MQC}
\begin{figure}[!ht]
	\begin{center}
	\includegraphics[scale=1]{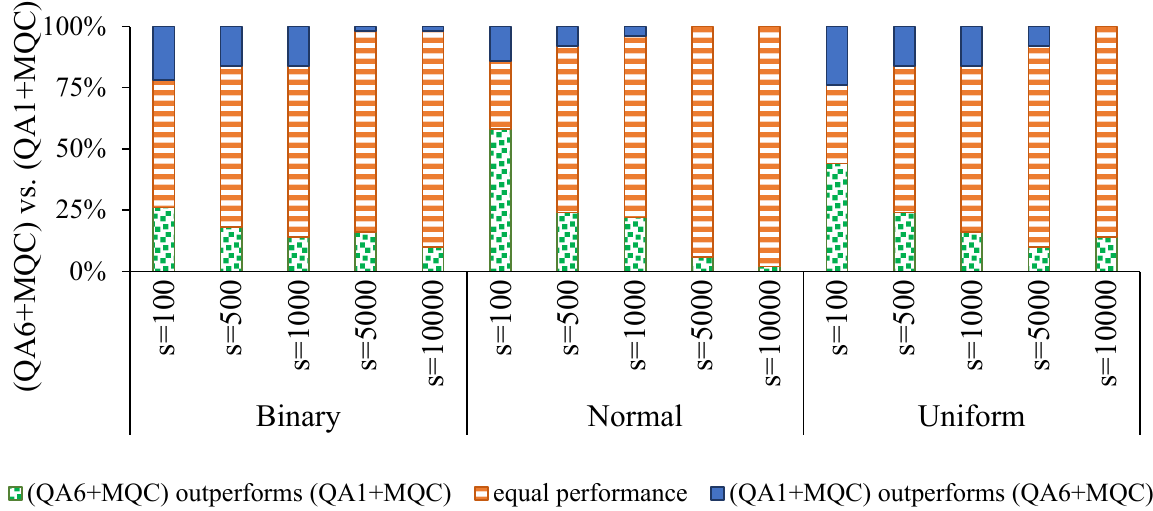} 
	\end{center}
	\vspace{-0.6cm}
	\caption{
	Comparing the performance of applying MQC to raw samples attained by a quantum annealer $(\text{QA}^1+MQC)$ with applying MQC to samples attained by the quantum annealing with five spin-reversal transforms, longer preparation time and optimization postprocessing $(\text{QA}^6+MQC)$ where $s$ denotes the number of samples. 
	}
	\label{fig:benchmarking_QA1+MQC_QA6+MQC}
\end{figure}

MQC is a postprocessing heuristic that iteratively reduces a set of samples to a smaller sample set whose energy values are lower than the previous iteration(s). 
Therefore, the performance of MQC mostly depends on samples that we obtain from the physical quantum annealers. 
When we repeat the annealing process on a D-Wave quantum annealer, successive measurements are correlated to each other due to limited preparation time. 
Hence, successive measurements generally form clusters of samples, i.e., groups of identical states. 
Since applying the Reduce procedure on identical input samples yields the same sample, early iterations of applying MQC on raw samples attained by the physical quantum annealers can become ineffective.
Figure \ref{fig:benchmarking_QA1+MQC_QA6+MQC} compares the performance of applying MQC to samples attained by standard quantum annealing (i.e., raw samples attained by the D-Wave quantum annealers) and enhanced quantum annealing (i.e., quantum annealing with spin-reversal transforms, longer inter-sample delay, and classical optimization postprocessing). 
For normal benchmark problems, results are statistically significant when we draw fewer than 5,000 samples. 
$p-$values of uniform and Binary problems are less than 0.05 only when we draw 10,000 and 5,000 samples, respectively. 

Figure \ref{fig:benchmarking_QA1+MQC_QA6+MQC} suggests that the quality of the input sample set can impact the performance of MQC. 
We propose the  \emph{randomized multi-qubit correction} (RMQC) scheme, presented in Algorithm \ref{algo:RMQC}, that repeats MQC on shuffled sample sets.
In RMQC, we repeat the MQC method $r$ times. 
We start with the raw sample set (similar to MQC) and then shuffle the sample set in each iteration. 
Finally, we apply MQC on $r$ samples (results from applying MQC on shuffled sample set) to obtain the final solution.
Note that when $r=1$ MQC and RMQC are identical. 
For $r>1$, RMQC is guaranteed to outperform MQC—albeit $r$ times more (classical) computation time/overhead.

\begin{algorithm}[!ht]
	\DontPrintSemicolon 
	\SetKwFunction{MQC}{MQC}
	\SetKwFunction{Append}{Append}
	\SetKwFunction{Shuffle}{Shuffle}
	\KwIn{$Z=\left[\mathbf{z}^1, \mathbf{z}^2, \dots, \mathbf{z}^n\right], r>0, \mathbf{h}$ and $J$}
	\KwOut{$\mathbf{z}^*$}
	$Z^* \gets \{\}$\;	
	\Do{$r>0$}{
		$r \gets r-1$\;	
		$Z^*.\Append{$\MQC{$Z$}{}$}{}$\;
		$Z \gets \Shuffle{$Z$}{}$\;	
	}	
	$\mathbf{z}^* \gets \MQC{$Z^*$}{}$\;
	\Return{$\mathbf{z}^*$}\;
	\caption{Randomized multi-qubit correction (RMQC).}
	\label{algo:RMQC}
\end{algorithm}

\begin{figure}[!ht]
	\begin{center}
	\includegraphics{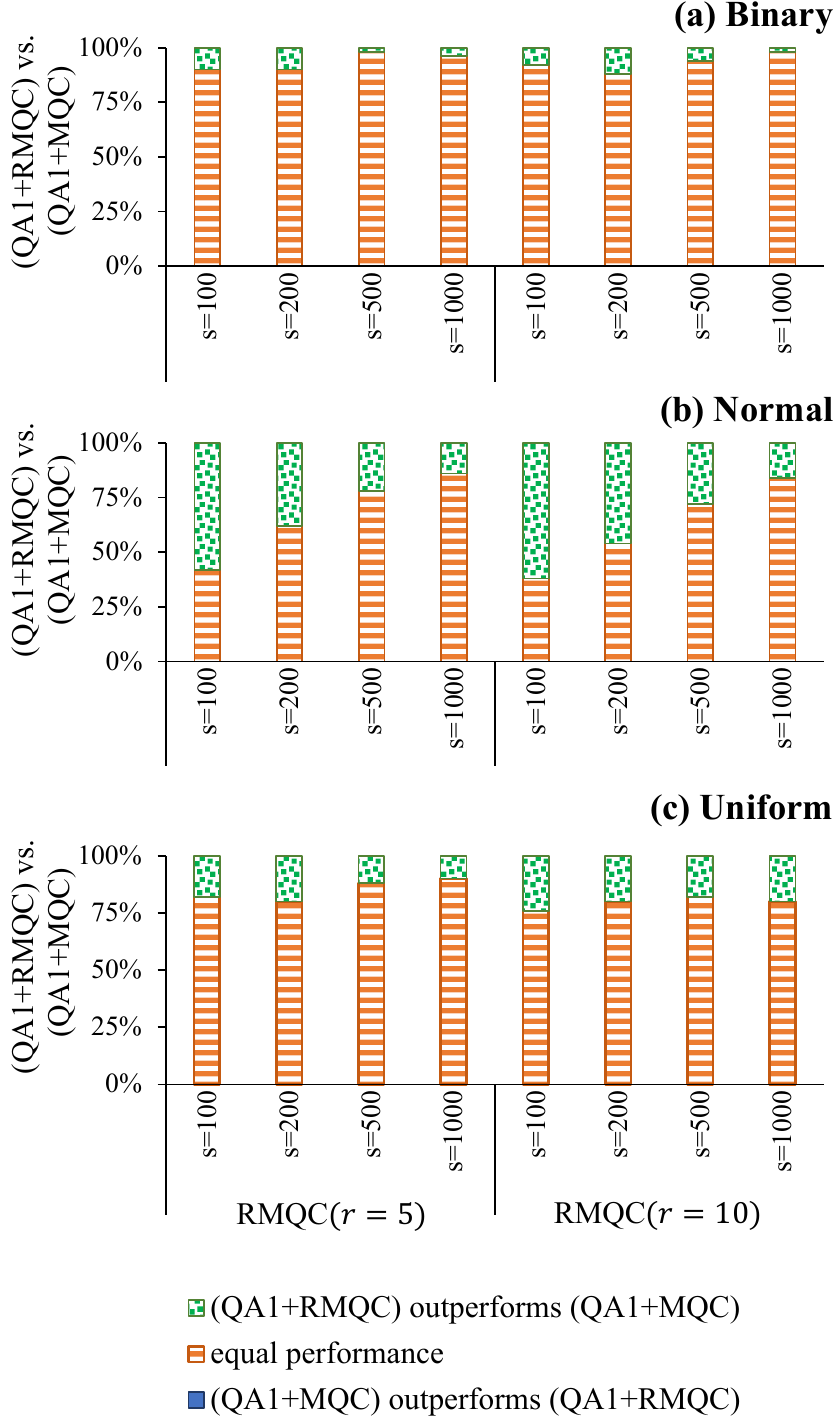} 
	\end{center}
	\vspace{-0.6cm}
	\caption{
	Performance comparison between MQC (denoted by $\text{QA}^1+MQC)$ and RMQC with $r=5$ and 10, denoted by $\text{QA}^1+RMQC$ $(s$ represents the number of samples).
	}
	\label{fig:benchmarking_QA1+MQC_QA1+RMQC51_101}
\end{figure}

Figure \ref{fig:benchmarking_QA1+MQC_QA1+RMQC51_101} illustrates the performance of RMQC for $r=5$ and $10$, and compares it with applying MQC to raw samples attained by the D-Wave quantum annealers.
For normal and uniform benchmark problems, results are always statistically significant (i.e., $p<0.05).$ 
More specifically, when $r=5,$ $p \in \left[{0.0078, 1.8626 \times 10^{-9}} \right]$ 
and $p \in \left[{0.0313, 0.001} \right]$ for normal and uniform benchmark problems, respectively. 
When $r-10,$  $p \in \left[{0.004, 4.6566 \times 10^{-10}} \right]$ 
and $p \in \left[{0.002, 0.0003} \right]$ for normal and uniform benchmark problems, respectively.
For Binary problems, however, results of MQC and RMQC are not statistically significant—for sufficient number of samples MQC and RMQC demonstrate similar performance on Binary benchmark problems. 

\section{Discussion}
Owing to various technological barriers such as diabatic transitions, thermal noise, and a vast range of control errors, quantum annealing in a real device (i.e., open-system and stoquastic) is necessarily susceptible to errors \cite{albash2018adiabatic,matsuura2019nested}. 
While several studies have proposed various error correction approaches for adiabatic quantum computers, most of them are not applicable to quantum annealers \cite{albash2018adiabatic}. 
Moreover, error correction techniques such as the nested quantum annealing correction method \cite{vinci2016nested,matsuura2019nested} use multiple physical qubits for representing logical qubits that notably reduce the capacity of current quantum annealers. 
Quantum annealers can draw many high-quality samples in near-constant time. In other words, the annealing time does not depend on the number of qubits. 
However, they generally fail to find the global minimum, specifically when the energy gap between the ground and the first excited state(s) is small.

While most studies try to recognize specific types of errors and mitigate/correct them, we show that one can recognize the pattern(s) of incorrect observations (or lower quality samples) and leverage it to improve the fidelity of the quantum annealers. 
In this sense, we view the open-system quantum annealing process as a Gibbs distribution sampler \cite{vinci2016nested} and exploit the neighborhood of the drawn samples to find a better solution. 
We first examined the impact of applying a local search heuristic, called \emph{single qubit correction} (SQC), on raw samples drawn by a D-Wave 2000Q quantum annealers.
Figure \ref{fig:benchmarking_QA1_QA1+SQC} reveals that applying SQC on raw samples for a given Ising Hamiltonian with normal and uniform coefficients always results in a sample with a lower energy value. 
In other words, none of the drawn samples for normal and uniform problems were a local optimum because SQC was able to perform a local search and find another sample with a lower energy value. 
On the other hand, roughly all drawn samples for binary problems were a local (or global) optimum, and applying SQC could not improve the quality of the attained samples.

There are two possibilities for this observation: 
(1) Ising Hamiltonians with binary (or discrete) coefficients are easier problems, and sampling with a D-Wave quantum annealer is very likely to result in the ground state; 
or (2) owing to the precision limitations (e.g., 8–9 bits precision on the D-Wave quantum annealers), the Ising Hamiltonian that is being minimized by a physical quantum annealer is different from the given Ising Hamiltonian (with double precision). 
Previous studies have mentioned that random Ising problems might not be hard-enough problems for quantum annealers (and even classical optimization techniques) \cite{katzgraber2014glassy,marshall2016practical}. 
In practice, nevertheless, we see that applying SQC/MQC can notably improve the fidelity of the D-Wave quantum annealers. 
More specifically, while for most random Binary problems applying SQC/MQC does not result in a better solution—the D-Wave quantum annealer can itself find the best (achievable) solution—applying MQC to random Normal and Uniform Ising Hamiltonians always results in a better solution. Hence, we can conclude that (in practice) easy to solve random Ising problems can be challenging for NISQ machines—we look at quantum annealers as the NISQ model of adiabatic quantum computers.
Indeed, not only does SQC exploits the neighborhood of an input sample for finding a sample with lower energy, but it can also remediate the precision limitations of physical quantum annealers. 

We extended SQC to introduce a novel postprocessing method, called \emph{multi-qubit correction} (MQC).
The first premise behind MQC is the idea that quantum annealers can draw high-quality samples from the ground state of the given problem Hamiltonian. 
In other words, all samples attained by a quantum annealer partially represent the ground state of the given problem Hamiltonian, although they can contain bits in error. 
The second premise is that we must simultaneously flip subsets of multiple bits to relax any excited state to a ground state. 
From another perspective, flipping bits individually (like how SQC tries to exploit the neighborhood of the measured samples) is (very) likely to result in a local optimum.

For every reduction, MQC takes two samples and bitwise compares them to determine which bits are the same and which are different. 
Although identical bits are more likely to be correct, we are not interested in them because we do not know whether they are correct. 
On the other hand, if a bit is different between the two samples, one of the samples has the correct bit value. 
The objective of MQC is to find groups of isolated bits such that simultaneously flipping them can yield a sample with a lower energy value.  
If there is only one isolated group, then flipping all the bits in the group will only change one of the two samples into the other. 

Figure \ref{fig:benchmarking_QA1+SQC_QA1+MQC} explains that applying MQC outperforms SQC. 
More specifically, for random normal and uniform Ising problems, applying MQC to raw samples always (i.e., in 100\% of the employed benchmark problems) results in a sample with lower energy than applying SQC to the same set of raw samples. 
On binary problems, nevertheless, MQC was able to outperform SQC in about 63\% of cases and they were a tie in approximately 37\% of the problems.
Figure \ref{fig:benchmarking_QAi_QA1+MQC} demonstrates that MQC outperforms recent software/hardware advances in the realm of quantum annealing—namely increasing the inter-sample delays and applying classical pre/post-processing methods that are available in Ocean SDK. 
More specifically, for normal and uniform random Ising problems, MQC always (100\% of the used benchmark problems) finds a sample with a lower energy value.

For binary problems, MQC almost results in a sample with a lower energy value. 
However, it is worth noting that applying spin-reversal transforms, longer inter-sample delay, and optimization postprocessing resulted in a better solution in about 2.3\%, 1\%, and  2\% of random binary problems, respectively. 
Besides, employing all enhancements (denoted by $\text{QA}^6)$ was able to outperform MQC in about 5\% of random binary problems. 
From an application viewpoint, near-term quantum processors provide a (noisy) distribution of the ground state(s). 
Hence, the results of the physical quantum annealers are not well reproducible, mostly due to thermal noise. 
Table \ref{tbl:benchmarking_robustness} reveals that applying MQC notably improves the robustness of the D-Wave quantum annealers. 

MQC might be considered related to sample persistence \cite{karimi2017boosting} or meta-heuristics (such as evolutionary algorithms and swarm intelligence techniques) \cite{shahamatnia2011adaptive}. 
MQC is a deterministic postprocessing method—i.e., for a fixed set of input samples, it will always yield the same solution—but most postprocessing schemes are stochastic. Hence, applying MQC (notably) improves the reproducibility of results attained by the quantum annealers. 
Moreover, most meta-heuristics require hyperparameter optimization (e.g., number of iterations, and probabilities of crossover and mutation in evolutionary algorithms); however, run-time and performance of MQC mainly depend on the size and quality of the input sample set, respectively.
More specifically, for ${n}$ samples, MQC performs $\log_2{n}$ iterations, and in iteration $i$ it compares $n/2^i$ pairs of samples. 

Successive measurements on the D-Wave quantum annealers are correlated to each other due to limited preparation time. 
From another perspective, successive measurements generally form clusters of samples (i.e., groups of identical states).
Consequently, technological barriers (e.g., the limited delay between successive reads and the thermal noise) can lessen the performance of MQC. 
Figure \ref{fig:benchmarking_QA1+MQC_QA6+MQC} shows that applying MQC to samples attained by enhanced quantum annealing (denoted by $\text{QA}^6)$—i.e., using five spin-reversal transforms, increasing the preparation time, and performing optimization postprocessing—mostly results in better solutions, compared to applying MQC on raw samples (denoted by $\text{QA}^1).$ 
It is worth highlighting that, nevertheless, increasing the number of reads/samples shrinks the gap between the performance of applying MQC to $\text{QA}^1$ to and $\text{QA}^6.$
In this sense, we extended MQC and introduced \emph{randomized MQC} (RMQC) that re-applies MQC on a shuffled sample set. 
RMQC is guaranteed to outperform MQC (for $r>1),$ albeit notably more (classical) computations.

\section{Method}
This paper presents a novel postprocessing method for quantum annealers, called \emph{multi-qubit correction} (MQC), that notably improves quantum annealers’ performance in terms of reproducibility of results and finding solutions with lower energy values.

\subsection{Quantum Hardware}
For all evaluations, we used a D-Wave 2000Q quantum annealer by D-Wave Systems Inc. (located at Burnaby, British Colombia, Canada). 
The annealing time for all experiments was 20 microseconds.

\subsection{Study Cases}
Generating random Hamiltonians is a common practice for benchmarking quantum annealers \cite{das2008colloquium,
pudenz2015quantum,
king2016degeneracy,
ayanzadeh2019quantum_assisted}. 
Hence, we generated three different types of Ising Hamiltonians as follows:
\begin{itemize}
	\item Binary—random Ising Hamiltonians whose linear and quadratic coefficients (denoted by $\mathbf{h}$ and ${J},$ respectively) were randomly drawn from $\{-1,+1\}$, based on a Bernoulli distribution with equal probabilities for -1 and +1;
	\item Uniform—random Ising Hamiltonians whose linear and quadratic coefficients (denoted by $\mathbf{h}$ and ${J},$ respectively) are (double-precision) uniform random numbers in $[-1,+1];$ 
	\item Normal—random Ising Hamiltonians whose linear and quadratic coefficients (denoted by $\mathbf{h}$ and ${J},$ respectively) are (double-precision) Normal random numbers that follow the standard Gaussian distribution, i.e., average and standard deviation are 0 and 1, respectively. 
\end{itemize}

For each problem type, we generated 50 instances (random problems) and adopted the finite-range Ising model, a.k.a. EA model (Edward—Anderson) \cite{das2008colloquium}, to generate benchmark problems. 
More specifically, all randomly generated benchmark problems were compatible with the D-Wave 2000Q quantum processor’s working graph—in Chimera topology, every qubit is connected to at most six other qubits.
Therefore, until the next maintenance that can change the quantum annealer’s working graph, one can directly execute them without embedding problems to a target graph. 

We remark that random Ising problems might not be hard-enough problems for quantum annealers (and even classical optimization techniques) \cite{katzgraber2014glassy}.
To demonstrate the quantum speed-up, one need to use proper benchmarks \cite{marshall2016practical}. 
It is worth highlighting that our objective in this study was not to show/claim quantum speed-up.

\subsection{Baselines}
Since executing a quantum machine instruction (QMI) on a physical quantum annealer is not guaranteed to achieve the ground state of the corresponding Ising Hamiltonian, even if we request many samples/reads, several studies have proposed software and hardware advancements to improve the performance of the quantum annealers. 
As an example, recent studies have revealed that using spin-reversal transforms (also known as gauge transforms)—i.e., flipping qubits randomly without altering the ground state of the original Ising Hamiltonian—can reduce analog errors of the quantum annealers \cite{pelofske2019optimizing}. 
Similarly, applying classical postprocessing heuristics on raw samples (attained by the quantum annealers) can result in samples with lower energy values \cite{borle2019post,golden2019pre}. 
Furthermore, when we submit a problem to a D-Wave quantum annealer, it is a common practice to request several samples/reads (i.e., up to 10,000 per QMI on the current D-Wave quantum processors). 
For every read (or measurement), the D-Wave QPU initializes all qubits and repeats the annealing process. 
Therefore, when we repeat the annealing process, successive measurements are correlated to each other due to limited preparation time. 
From another perspective, successive measurements generally form clusters of samples (i.e., groups of identical states). 
Hence, increasing the preparation time can reduce the inter-sample correlations. 
In this study, we used the following arrangements for evaluating the performance of the proposed postprocessing methods: 
\begin{itemize}
	\item $\text{QA}^1$—raw samples attained by a D-Wave quantum annealer;
	\item $\text{QA}^2$—applies five spin-reversal transforms on $\text{QA}^1;$
	\item $\text{QA}^3$—puts a longer delay between successive reads/samplings to reduce the sample-to-sample correlation, albeit longer run-time;
	\item $\text{QA}^4$—performs the optimization postprocessing, available from the D-Wave’s Ocean SDK, to all raw samples ($\text{QA}^1);$
	\item $\text{QA}^5$— performs the sampling postprocessing, available from the D-Wave’s Ocean SDK, to all raw samples ($\text{QA}^1);$
	\item $\text{QA}^6$—applies five  spin-reversal transforms, puts a longer delay between successive reads, and performs the optimization postprocessing to raw samples (attained by $\text{QA}^1).$ 
\end{itemize}

\subsection{Evaluations}
To compare MQC with baseline methods, we count how many times the \emph{expected to win} method outperforms the baseline (i.e., finds a sample with lower energy value), ignoring cases where both methods can find an identical solution (in terms of energy value). 
Since case studies used here are all random problems, of interest is how statistically significant such results are. 
Hence, we performed hypothesis testing. 
Our null hypothesis in all evaluations was that the \emph{expected to win} method and the baseline are not different, and the alternate hypothesis was that the \emph{expected to win} method outperforms the baseline (i.e., can always find a sample with lower energy value). 
To this end, we calculate the $p-$value as:
\begin{equation}
	\label{eqn:p-value}
	p=\frac{1}{2^{n_b+n_w}}\sum^{n_b+n_w}_{k=n_b} \binom{n_b+n_w}{k} \, .
\end{equation}
where $n_b$ represents the number of times that the \emph{expected to win} method was able to outperform the baseline and $n_w$ is the number of times that the baseline was able to outperform the \emph{expected to win} method. 
This represents the probability of the null hypothesis is true. 
By convention, $p < 0.05$ disproves the null hypothesis (i.e., the \emph{expected to win} indeed outperforms the baseline) and is considered to be statistically significant, 
and $0.05 \leq p \leq 0.95$ indicates that results are not statistically significant.  
While $p \geq 0.95$ also disprove the null hypothesis (i.e., results are statistically significant), it also rejects the alternate hypothesis and indicates that the expected winner has been chosen incorrectly \cite{chen2021performance}.

\section*{Acknowledgements}
This research was carried out when the first author was a student member and research faculty member at the University of Maryland, Baltimore County. He is currently a postdoctoral fellow at the Georgia Institute of Technology.
This research has been supported by NASA grant (\#NNH16ZDA001N-AIST 16-0091), NIH-NIGMS Initiative for Maximizing Student Development Grant (2 R25-GM55036), and the Google Lime scholarship. We would like to thank the D-Wave Systems management team for granting access to the D-Wave 2000Q quantum processor.

\section*{Author contributions statement}
R.A. and J.D. conceived algorithms, and R.A. carried out implementations. M.H. and T.F. performed analysis. 

\section*{Additional information}
Authors declare no competing interests.

\bibliography{bibliography/quantum,bibliography/SAT,bibliography/AI_ML,bibliography/mypublications,bibliography/CS,bibliography/HPC}

\end{document}